\documentclass[aps,prd,showpacs,nofootinbib,floats,floatfix,preprintnumbers,groupedaddress,twocolumn]{revtex4}
\usepackage{comment}
\usepackage{cancel}
\usepackage{graphicx}
\usepackage{amsmath, mathrsfs}
\usepackage{float}
\usepackage{subfigure}
\usepackage[usenames]{color}
\usepackage{amssymb}
\usepackage{bbm}
\usepackage{csquotes}

\usepackage{tikz}
\usetikzlibrary{positioning,decorations.pathmorphing}

\usepackage{epstopdf}
\newcommand{\bea}{\begin{aligned}}
\newcommand{\eea}{\end{aligned}}
\newcommand{\beq}{\begin{equation}}
\newcommand{\eeq}{\end{equation}}

\newcommand{\bse}{\begin{subequations}}
\newcommand{\ese}{\end{subequations}}

\usepackage{hyperref}
\hypersetup{
     colorlinks   = true,
     citecolor    = red,
     linkcolor    = blue,
     urlcolor     = blue,
}

\newcommand{\bmm}{\begin{multline}}
\newcommand{\emm}{\end{multline}}

\begin{document}
\title{
%Non-Planckian/Modified Hawking spectrum in thermal bath and PBH cosmology\\
%Thermal correction to Hawking radiation and its application to PBHs\\
%Thermal corrections lower black hole lifetime: implications for primordial black holes\\
Black holes in thermal bath live shorter: Implications for primordial black holes}
%%%%%%%%%%%%%%%%%%%
\author{Jitumani Kalita}
\email{E-mail: k.jitumani@iitg.ac.in}
\affiliation{Department of Physics, Indian Institute of Technology, Guwahati, 
Assam, India}
%%%%%%%%%%%%%%%
\author{Debaprasad Maity}
\email{E-mail: debu@iitg.ac.in}
\affiliation{Department of Physics, Indian Institute of Technology, Guwahati, 
Assam, India}
%%%%%%%%%%%%%%%%%
\author{Ayan Chatterjee}
\email{E-mail: ayan.theory@gmail.com}
\affiliation{Department of Physics \& Astronomical Science, Central University of Himachal Pradesh, Dharamshala-
176215, India.}
%%%%%%%%%%%%%%%%%%%%%%%%%%
\pagenumbering{arabic}
%\maketit
\renewcommand{\thesection}{\arabic{section}}
%%%%%%%%%%%%%%%%%%%%%
\begin{abstract}
Hawking radiation from a non- extremal black hole is known to be approximately Planckian. 
The thermal spectrum receives multiple corrections including
greybody factors and due to kinematical restrictions on the infrared and ultraviolet frequencies.
We show that another significant correction
to the spectrum arises if the black hole is assumed to live in a thermal bath and the emitted radiation gets thermalised at the bath temperature. This modification reshapes the thermal spectrum, and leads to appreciable
deviation from standard results including modification in the decay rate of black holes.
We argue 
that this altered decay rate has significance for 
%phenomenological impact in 
%the context of 
%primordial black holesin
%early universe 
cosmology and, in a realistic setting, show 
that it alters 
the life time of primordial black holes (PBHs) in the early universe. In particular, the very light PBHs formed right after the end of inflation decay faster which may have interesting phenomenological implications.  
\end{abstract}
%%%%%%%%%%%%%%%%%%%%%%

\maketitle
\newpage

The horizon of a non- extremal 
black hole acquires a temperature
$\kappa/2\pi$ due to quantum phenomena (using $\hbar=c=k_{B}=1$ units), where
$\kappa$ is the surface gravity of the horizon \cite{Hawking:1975vcx}. Due to this thermality,
an observer at $\mathscr{I}^{+}$
receives a steady flux of Hawking radiation of \emph{approximately} Planckian nature.
%For the observers at $\mathscr{I}^{+}$,
%the frequency distribution and temperature of the emitted flux is a direct consequence of 
%the presence of a trapped region. 
%More precisely, 
The thermal flux arises due to the presence of trapped surfaces in the black hole region and, the amount of emitted outgoing positive energy flux through $\mathscr{I}^{+}$ is related to the 
decrease of horizon radius \cite{Chatterjee:2012um}. %whereas, the thermality is attributed to tracing over the causally disconnected trapped
%region \cite{Israel:1976ur}. 
This is the scenario according to standard formulation of black hole radiance,
which has been confirmed through alternate techniques 
and, has unambiguously revealed the correctness of (observer dependent)
thermal nature of the quantum vacuum \cite{Israel:1976ur, Hartle:1976tp, Gibbons:1976pt, Wald:1995yp}. 
In conjunction with the laws of black hole mechanics, 
this phenomena has definitely established that horizons obey the laws of thermodynamics. 
%On one hand, this thermal property substantiates 
%the inherent quantum nature of horizons, but on the other, highlights fundamental inconsistencies, collectively dubbed as the information loss puzzle \cite{Hawking:1976ra,Harlow:2014yka,Mathur:2009hf,Raju:2020smc,Perez:2017cmj}. 
%The guiding avenues in quantum gravity, those based on 
%the AdS- CFT or Loop Quantum Gravity (LQG), suggest that the resolution of this paradox lies in a complete theory of quantum gravity \cite{Raju:2020smc,Perez:2017cmj}.
However, this standard formulation of Hawking radiation described above, also suffers from a major drawback - it assumes that fields at $\mathscr{I}^{-}$ are at zero temperature whereas, it is natural to expect that initially, the quantum fields are in equilibrium at a finite non-zero temperature. Consequently, it is essential to include modifications arising due to quantised thermal fields into the radiation formula. In this paper, we show that such a reasonable correction leads to substantial modification in the flux formula and, in particular, also incorporates temperature of the thermal field as a parameter. To show that the modification is not miniscule, and as a concrete application, we demonstrate that in the early universe, the decay rate of PBHs is appreciably increased, resulting in their shorter lifetimes. \\

 The shape of Hawking flux is approximately Planckian, although several corrections alter its spectrum mildly. Such alterations are natural in light of the built-in approximations in the original calculation. The first of these is the greybody factor, a consequence of the backscattering of Hawking flux due to the gravitational field of the horizon \cite{Page:1976df, Page:1976ki, Bekenstein:1977mv,Boonserm:2008zg}. This frequency dependent factor measures the absorption cross-section of the black hole and is responsible for the suppression of flux at infinity. A second correction arises as a result 
 of the fact that during this dynamical process of flux emission, the zeroth law remains approximately valid, $\kappa$ varies slowly \cite{Hawking:1975vcx,Chatterjee:2012um,Barcelo:2010pj, Visser:2001kq}. This assumption constraints the Hawking flux to  include only the frequencies $\omega\ge \sqrt{\dot{\kappa}}$. The spectrum also receives a third correction, an ultraviolet cut-off to its frequency. This is due to the kinematical phase- space constraints that the frequency of emitted particle cannot exceed the available ADM mass of the spacetime, $\omega\le M_{ADM}$ \cite{Visser:2014ypa,Parikh:1999mf}. 
  Are further rectifications to the black hole spectrum possible over and above these corrections? As mentioned earlier, 
 the answer is in the affirmative: There does exists a fundamental modification which has hereto remained unexplored, and pertains to the corrections that the flux suffers if the emitted particles from the black hole living in a thermal bath is assumed to be thermalized.\\

To formulate the problem, consider a massless quantum matter field (either scalars or spin-1/2 fermions) in equilibrium at a non- zero temperature $T_{b}=\beta^{-1}$ on $\mathscr{I}^{-}$. On $\mathscr{I}^{-}$, one may also introduce a (thermal) vacuum, defined to be a state such that the ensemble average of any observable in thermal equilibrium at temperature $T_{b}$ may be obtained from its expectation value in this thermal state \cite{Takahashi:1996zn, Das_book}. To an inertial observer at $\mathscr{I}^{-}$, this vacuum state is thermal, with each mode being Planckian at $\beta$.  
%Thus, for any operator $A$ associated to the quantum system, the expectation value in this thermal state is given by: 
%$ \langle A \rangle _{_{\beta}}=Z(\beta)^{-1}\,\text{Tr}\,(e^{-\beta H}A )$, where $Z(\beta)=\text{Tr}(e^{-\beta H})$ is the partition function of the system, and $H$ denotes Hamiltonian. 
If this thermalised quantum field is evolved in the collapsing geometry of a spherical black hole, 
%For simplicity, it will be assumed that the quantum field has no interaction with the collapsing matter forming the black hole scatters from the black hole to reach observers at $\mathscr{I}^{+}$. 
we demonstrate that, to observers at $\mathscr{I}^{+}$,
the thermal vacuum state corresponding to the in modes remains thermally populated, but the energy distribution is altered and is now manifestly a function of both temperatures, 
the initial temperature of the thermal field
and the black hole temperature. So, although the flux remains thermal to future observers, the energy spectrum is not Planckian anymore.
%Furthermore, as might be expected, the final spectrum at infinity is commensurate in its dependence on either temperatures.
This modification is
substantial enough to warrant a 
thorough scrutiny into its impact on various physical phenomena. In particular, we show that 
the corrected flux formula enhances the decay rates of black holes, and such enhancement may have potential implication for the evolution of a black hole living in thermal bath. In the context of early universe cosmology, the scenario formulated here presents itself naturally- we indeed have black hole-thermal bath systems wherein black holes are produced from large primordial fluctuations. Utilizing the modified decay rate formula we also investigate the dependence of mass of PBHs on their lifetimes since it is understood that such changes in the dynamics of decay rates may have significant effect in the cosmology of very universe. This paper is therefore expected to add to the ongoing effort to understand the formation, evolution and impact of PBHs in a cosmological setting \cite{Chapline:1975ojl, Carr:2016drx, Carr:2020gox, Carr:2009jm}. \\

%existing Big-Bang Neucleosynthesis (BBN) 

To determine the evolution of quantised thermal fields on the black hole background, we make use of the thermofield double (TFD) state formalism \cite{Takahashi:1996zn, Das_book}. This formalism has been used to understand properties of thermalised matter fields, to define thermal states of black holes \cite{Israel:1976ur}, and to describe holographic duals of black holes in the AdS/CFT \cite{Maldacena}. It is also expected to play a fundamental role in the resolution of information loss puzzle in gauge- gravity duality. 
%\subsubsection*{Finite temperature Hawking radiation for spin $0$}
%\textbf{Thermofield-double formalism}
%\noindent
The TFD formalism doubles the degrees of freedom of the system by supplementing
the standard Fock space $\mathscr{F}$ with another, made-up Fock space $\mathscr{\Bar{F}}$, also called the dual, to create $\mathscr{F}\otimes \mathscr{\Bar{F}}$.
%Since these two Fock spaces are defined in different spacetimes, the corresponding fields are uncoupled in the Lagrangian, although they remain quantum mechanically correlated. 
On $\mathscr{F}\otimes \mathscr{\Bar{F}}$, there exists a temperature dependent vacuum, called the TFD state, such that the ensemble average of any observable can be obtained by calculating the expectation value in this thermal state. More precisely, if $\vert 0, \beta \rangle $ is the TFD state,
\begin{equation}\label{tfd_state}
\vert 0, \beta \rangle = \frac{1}{\sqrt{Z(\beta)}}\, \sum_n\, e^{-\frac{\beta E_n}{2} }\,\vert n\rangle \otimes{\vert\tilde n\rangle}, 
\end{equation}
where $\vert n\rangle \otimes{\vert\tilde n\rangle} \in (\mathscr{F}\otimes \mathscr{\Bar{F}})$, are the energy eigenstates
associated with two identical entangled quantum systems and $\beta$ is the inverse temperature of the system. Then, for any operator $\hat{A}$, associated to the actual quantum system, $\langle \hat{A} \rangle _{_{\beta}}=Z(\beta)^{-1}\,\text{Tr}\,(e^{-\beta \hat{H}}\hat{A})$,
%%%%%%%%%%%%%%%
where expectation value is on the TFD state, and $Z(\beta)=\text{Tr}(e^{-\beta \hat{H}})$ is the partition function of the system with $\hat{H}$ representing Hamiltonian of the original system. In short,
the TFD state is a pure state in $\mathscr{F}\otimes \mathscr{\Bar{F}}$, but is thermal if the augmented system is traced out. In the context of quantisation of thermal fields, the imaginary system $\mathscr{\Bar{F}}$ is for book-keeping purpose, whereas in the context of black holes, its nature is physical, since one may associate the augmented system to the part of spacetime beyond the horizon and the trapped region of the extended Kruskal spacetime. Thus, in this sense, the TFD state is attributed
a physical meaning for a
black hole spacetime.\\

To develop the formalism, consider a massless scalar field
$\phi(x)$, in equilibrium with
thermal bath of inverse temperature $\beta$ on $\mathscr{I}^{-}$.
The field decomposes in terms of ingoing $\lbrace g_{\omega} \rbrace$, and outgoing modes $\lbrace h_{\omega} \rbrace$,
with their corresponding operators $\hat{B}^{\pm}_{\omega}$ and $\hat{C}^{\pm}_{\omega}$,
\begin{equation}
{\phi}(x)=\int_0^{\infty} d\omega \left[g_{\omega}{B}^{-}_{\omega}+g^{\ast}_{\omega}{B}^{+}_{\omega}+h_{\omega}{C}^-_{\omega}+h^{\ast}_{\omega}{C}^+_{\omega} \right].
\end{equation}
In the following, we concentrate only on the ingoing modes, a similar analysis follows for the outging modes too. 
If $\vert 0\rangle$ denotes the vacuum state, then  ${B}^{-}_{\omega}\vert 0\rangle=0$. The single particle states $\vert n_{\omega}\rangle$ of 
the Fock space $\mathscr{F}$ are eigenstates of the number operator ${N}^{(B)}_{\omega}={B}^+_{\omega}{B}^-_{\omega}$, such that  $
{N}^{(B)}_{\omega}\vert n_{\omega}\rangle=n_{\omega}\vert n_{\omega}\rangle $. To implement the TFD formalism, we augment the Fock space $\tilde{\mathscr{F}}$ to the existing space. The states of $\tilde{\mathscr{F}}$ and operators acting on it are denoted by the tilde symbol. For example,  ${\vert \tilde{n}_{\omega}\rangle}\in \tilde{\mathscr{F}}$, and $\tilde{N}^{(\tilde{B})}_{\omega}=\tilde{{B}}^{+}_{\omega}\tilde{{B}}^{-}_{\omega}$ is the corresponding number operator. Naturally, the product state is denoted by  $\vert n_{\omega},\tilde{n}_{\omega}\rangle=\vert n_{\omega}\rangle\otimes \vert \tilde{n}_{\omega}\rangle$.
The associated TFD state  is obtained from \eqref{tfd_state}.
%the  associated TFD state, $\vert 0, \beta %\rangle$ for the thermal scalar field would be,
%
%\begin{equation}
 %   \vert 0, \beta \rangle =Z^{-\frac{1}%{2}}\sum_{n_{\omega}} e^{-\int_{\omega}d\omega %N^{(B)}_{\omega}\frac{\beta\omega}{2}} \vert %n_{\omega},\tilde{n}_{\omega}\rangle.
%\end{equation}
%
This is a correct choice of state since, a simple calculation reveals
\begin{equation}\label{thermal_scalar1}
    \langle 0,\beta\vert{N}^{(B)}_\omega\vert 0,\beta\rangle =\frac{1}{e^{\beta\omega}-1},
\end{equation}
which is the appropriate thermal spectrum for each mode $\omega$ in thermal equilibrium at $\beta$. \\

The construction of TFD state and the action of operators is often complicated and therefore, it is simple to rephrase them in terms of \emph{thermal operators} acting on zero temperature vacuum. Let, $U(\theta)$ be an unitary operator which transforms the vacuum $\vert 0, \tilde{0}\rangle$ to the thermal vacuum
$\vert 0, \beta \rangle$, such that $U(\theta) \vert 0, \tilde{0}\rangle=\vert 0, \beta \rangle $. Then, any zero temperature operator $A$ may be reexpressed in terms of its thermal counterpart $A(\beta)$, using $A(\beta)=U(\theta)\, A\,\, U^{\dagger}(\theta)$. In particular, given a specific unitary operator
\begin{equation}
    U(\theta)=\exp \left[ {-\int_{\omega}d\omega\,\theta_{\omega}(\beta)\left(\tilde{{B}}^-_{\omega}{B}^-_{\omega}-{B}^+_{\omega}\tilde{{B}}^+_{\omega}  \right)} \right], 
\end{equation}
where $\theta_{\omega}(\beta)$ is given by 
\begin{equation}
    \cosh\theta_{\omega}(\beta)=\frac{1}{\sqrt{1-e^{-\beta\omega}}}~~;~~\sinh\theta_{\omega}(\beta)=\frac{e^{-\frac{\beta\omega}{2}}}{\sqrt{1-e^{-\beta\omega}}},
\end{equation}
the thermal annihilation and creation operators becomes 
\begin{eqnarray}\label{thermal_creation_operator}
{B}^{\pm}_{\omega}(\beta) &=&{B}^{\pm}_{\omega}  \cosh\theta_{\omega}(\beta)-\tilde{{B}}^{\pm}_{\omega}\sinh\theta_{\omega}(\beta).
\end{eqnarray}
To verify that the thermal operators give the correct expression on thermal vacuum, consider the action of number operator ${N}^{(B)}_{\omega}(\beta)= {B}^{+}_{\omega}(\beta){B}^{-}_{\omega}(\beta) $ in the vacuum $\vert 0,\tilde{0} \rangle$. This gives precisely the same spectrum, as in eqn. \eqref{thermal_scalar1} as required. Thus, the introduction of thermal operators allows us to shift focus from the TFD state $\vert 0, \beta\rangle$ to the product vacuum state $\vert 0,\tilde{0} \rangle$. This transition will play an important role towards introduction of thermal creation and annihilation operators in eqn. \eqref{thermal_creation_operator} and paves way for its application towards the modification of the Hawking spectrum.\\

In view of this development, the massless scalar field in thermal equilibrium at $\beta$ at $\mathscr{I}^-$, has 
a field decomposition in terms of \emph{thermal} operators:
\begin{equation}
\phi_{\beta}(x)=\int_0^{\infty} d\omega \left[ \, f_{\omega}\,{a}^-_{\omega}(\beta)+f^{\ast}_{\omega}\,{a}^+_{\omega}(\beta) \, \right],
\end{equation}
where, $\lbrace f_{\omega}\rbrace $ are the set of positive frequency ingoing modes. Together with   $\lbrace f^{*}_{\omega}\rbrace$, it forms a complete orthonormal set on $\mathscr{I}^-$ satisfying:
\begin{eqnarray}\label{ortho_normal}
    (f_{\omega}, f_{\omega^{\prime}} ) &=&\delta (\omega-\omega^{\prime}), ~  (
    f^{*}_{\omega}, f^{*}_{\omega^{\prime}} )=-\delta (\omega-\omega^{\prime}),\nonumber\\
     (f_{\omega}, f^{*}_{\omega^{\prime}} )&=&0.
\end{eqnarray}
It is easy to check that on $\mathscr{I}^-$, the expectation value of the thermal number operator $N^{(a)}_{\omega}(\beta)=a_{\omega}^+(\beta)\, a_{\omega}^-(\beta)$  on the state $\vert 0,\tilde{0} \rangle_{-}$ gives the correct thermal spectrum, as in eqn. \eqref{thermal_scalar1}.\\
%
%\begin{equation}
%{}_-\langle 0,\tilde{0} \vert N^{(a)}_{\omega}(\beta)\vert 0,\tilde{0} \rangle_-=\frac{1}{e^{\beta\omega}-1}.
%\end{equation}
%

Next, the quantum field on $\mathscr{I}^+$ can be written in terms of ingoing and outgoing modes:
\begin{equation}
{\phi}_{\beta}(x)=\int_0^{\infty} d\omega \left[p_{\omega}{b}^-_{\omega}(\beta)+p^{\ast}_{\omega}{b}^+_{\omega}(\beta)+q_{\omega}{c}^-_{\omega}(\beta)+q^{\ast}_{\omega}{c}^+_{\omega}(\beta) \right]
\end{equation}
where, $\lbrace p_{\omega}\rbrace$ and $\lbrace q_{\omega}\rbrace$ are the modes related to $\mathscr{I}^+$ and the future horizon $\mathcal{H}^{+}$ respectively and ${b}^{\pm}_{\omega}(\beta)$ and ${c}^{\pm}_{\omega}(\beta)$ are the corresponding thermal operators, related to their zero temperature counterparts through eqn.
\eqref{thermal_creation_operator}. Note that the outgoing modes $\lbrace p_{\omega}\rbrace$ may be expressed as a linear combination of the complete mode basis $\lbrace f_{\omega},f^{\ast}_{\omega}\rbrace$,
\begin{align}
p_{\omega} &=\int_0^{\infty}d\omega^{\prime}\left(\alpha_{\omega\omega^{\prime}}f_{\omega^{\prime}}+\beta_{\omega\omega^{\prime}}f^{\ast}_{\omega^{\prime}} \right).
%f_{\omega^{\prime}} &=\int_0^{\infty}d\omega \left(\alpha^{\ast}_{\omega\omega^{\prime}}p_{\omega}-\beta_{\omega\omega^{\prime}}p^{\ast}_{\omega} \right).
\end{align}
where $\alpha_{\omega\omega^{\prime}} =\left(p_{\omega},f_{\omega^{\prime}} \right)$, and $~~\beta_{\omega\omega^{\prime}} =-\left(p_{\omega},f^{\ast}_{\omega^{\prime}} \right)$.
%The orthonormality conditions of $\lbrace f_{\omega}\rbrace$ in eqn. \eqref{ortho_normal} leads to evaluation of 
%the coefficients $\alpha_{\omega\omega^{\prime}}$ and $\beta_{\omega\omega^{\prime}}$:
%
%\begin{align}
%\alpha_{\omega\omega^{\prime}} =\left(p_{\omega},f_{\omega^{\prime}} \right), ~~\beta_{\omega\omega^{\prime}} =-\left(p_{\omega},f^{\ast}_{\omega^{\prime}} \right). \label{1.3.8}
%\end{align}
%
The thermal operator $a_{\omega}(\beta)$ and $b_{\omega}(\beta)$ are related:
\begin{align}
{b}^{+}_{\omega}(\beta) &=\int_0^{\infty}d\omega^{\prime}\left[\alpha_{\omega\omega^{\prime}}{a}^{+}_{\omega^{\prime}}(\beta)-\beta_{\omega\omega^{\prime}}a^{-}_{\omega^{\prime}} (\beta) \right]
\end{align}
%Compairing both sides, the operators are related as
%\begin{align}
%\hat{b}^-_{\omega} &=\int_0^{\infty}d\omega^{\prime}\left(\alpha^{\ast}_{\omega\omega^{\prime}}{a}^{-}_{\omega^{\prime}}(\beta)-\beta^{\ast}_{\omega\omega^{\prime}}\hat{a}^{+}_{\omega^{\prime}} \right) 
%\hat{b}^+_{\omega} &=\int_0^{\infty}d\omega^{\prime}\left(\alpha_{\omega\omega^{\prime}}\hat{a}^+_{\omega^{\prime}}-\beta_{\omega\omega^{\prime}}\hat{a}^{-}_{\omega^{\prime}} \right)
%\end{align}
Naturally, the thermal number operator of $\mathscr{I}^+$, given by $N^{(b)}_{\omega}(\beta)={b}_{\omega}^+(\beta){b}_{\omega}^-(\beta)$, acting on the $\mathscr{I}^{+}$ vacuum $\vert 0,\tilde{0} \rangle_{+}$ gives the thermal spectrum corresponding to the temperature $\beta$. But if $N^{(b)}_{\omega}(\beta)$ is evaluated on the $ \vert 0,\tilde{0} \rangle_{-}$, it gives corrections to the Hawking radiation.
To evaluate this, assume that 
a fraction $\Gamma_{\omega}$ of $ p_{\omega}$ enters the collapsing matter configuration and is affected by the gravitational potential of the black hole. The fraction $\Gamma_{\omega}$ is
given by:
\begin{align}\label{2.3.13}
\int_0^{\infty} d\omega^{\prime}\left( \,\vert\alpha_{\omega\omega^{\prime}}\vert^{2}-\vert\beta_{\omega\omega^{\prime}}\vert^{2} \,\right)=\Gamma_{\omega}\delta(0)
\end{align}
Using the eikonal approximation, the coefficients $\alpha_{\omega\omega^{\prime}}$ and $\beta_{\omega\omega^{\prime}}$ are related by \cite{Hawking:1975vcx},
\begin{equation}
\vert \beta_{\omega\omega^{\prime}}\vert ^2=e^{-\frac{2\pi\omega}{\kappa}}\vert \alpha_{\omega\omega^{\prime}}\vert^2 .
\end{equation}
Therefore, the particle production at finite temperature is given by the value of $\mathscr{N}_{\omega}={}_{-}\langle 0,\tilde{0}\vert N^{(b)}_{\omega}(\beta)\vert 0,\tilde{0} \rangle_{-}$:
\begin{align}
\mathscr{N}_{\omega} &=\cosh^2 \theta_{\omega}\int_0^{\infty}d\omega^{\prime} \vert \beta^{(2)}_{\omega\omega^{\prime}}\vert^2+\sinh^2 \theta_{\omega}\int_0^{\infty}d\omega^{\prime} \vert \alpha^{(2)}_{\omega\omega^{\prime}}\vert^2 \notag \\
&= \frac{\Gamma_{\omega}}{e^{\frac{2\pi\omega}{\kappa}}-1}\left[1+ \frac{e^{\frac{2\pi\omega}{\kappa}}+1}{e^{\beta\omega}-1} \right]\delta(0).
\end{align}
Therefore, the number density of outgoing scalar particles in each mode is distributed in accordance with a modified 
spectrum:
\begin{equation}
n^{\rm Scalar}_{\omega}=\frac{\Gamma_{\omega}}{e^{\frac{2\pi\omega}{\kappa}}-1}\left[1+ \frac{e^{\frac{2\pi\omega}{\kappa}}+1}{e^{\beta\omega}-1} \right].\label{scalar_ndensity}
\end{equation}
For spin $1/2$-fermions, the number density can also be obtained and the expression is:
\begin{equation}
n^{\rm Fermion}_{\omega}=\frac{\Gamma_{\omega}}{e^{\frac{2\pi\omega}{\kappa}}+ 1}\left[1+ \frac{e^{\frac{2\pi\omega}{\kappa}}- 1}{e^{\beta\omega}+ 1} \right].\label{fermion_ndensity}
\end{equation}
Clearly, the equations reproduce two important liming spectrum. They approach the original Hawking spectrum at zero bath temperatures $(\beta\rightarrow \infty)$. In the zero black hole temperature limit $T_{\rm BH} \rightarrow 0$, they approach to pure background thermal bath spectrum $n_w\rightarrow n^{\beta}_w = \Gamma_w/(e^{\beta\omega}\pm1)$. Note that in their cosmological evolution, the BH mass evolution therefore would be governed by the subtracted spectrum $n_w\rightarrow n_w - n_w^{\beta}$.
%In the high temperature limit ($\beta\omega<<1$), the above expression reduces to:
%
%\begin{eqnarray}
%&&  n^{\rm Scalar}_{\omega} \simeq \frac{\Gamma_{\omega}}{e^{\frac{2\pi\omega}{\kappa}}+ 1} \frac{1}{\beta \omega} \\
%&& n^{\rm Fermion}_{\omega}=\frac{\Gamma_{\omega}}{e^{\frac{2\pi\omega}{\kappa}}+ 1}e^{\frac{2\pi\omega}{\kappa}}  
%\end{eqnarray}
%
These are the main result of this paper: it represents a non- Planckian correction to the original Hawking formula due to thermalised quantum fields. Apart from the fact that the improved number density in \eqref{scalar_ndensity} or \eqref{fermion_ndensity} appears to be interesting in their own right, and may be of fundamental importance to warrant further investigation, such finite temperature modification may cause revision of our perspectives on cosmological evolution.
Therefore, in the following, we apply this modified thermal spectrum and note its ramifications, particularly for the early universe cosmology, where the role of PBHs and their decay modes are of high significance. 
%effort, where they are assumed to be  formed surrounded by high temperature thermal bath\cite{jgjg}. This is precisely where
\\

Assume that a black hole of mass $M$ is in a thermal bath of temperature $T_b$. Due to the modified Hawking spectrum given above in eqns. \eqref{scalar_ndensity} and \eqref{fermion_ndensity}, the BH decay dynamics will be governed by the following particle emission rate \cite{Page:1976df, Cheek:2021odj} 
%In the high temperature limit
%\begin{equation}
%n_{\omega}=\frac{e^{\frac{2\pi\omega}{\kappa}}+1}{e^{\frac{2\pi\omega}{\kappa}}-1} \frac{T\Gamma_{\omega}}{\omega} 
%\end{equation}
%\section{Rate of decay of BH}
%
\begin{equation}
    \frac{d^2N_{i}}{dp dt}=\frac{g_i}{2\pi} \frac{p}{\omega} ({n}_{\omega} - n_w^{\beta}) ,
\end{equation}
where $p=\sqrt{\omega^2-\mu_i^2}$ is the momentum of the emitted  particle, and $(\mu_i,g_i)$ denote the mass and internal degrees of freedom of the $i$-th particle species, respectively. Integrating over all momenta and summing over all the SM particles one can write the BH decay rate as
\begin{equation}\label{Page_eqn1}
    \frac{dM}{dt}=-\sum_{i}\int_0^{\infty}\omega \frac{d^2 N_i}{dp dt}dp=-\frac{M^4_{p}}{M^2}\sum_{i}g_i \epsilon_i .
\end{equation}
%Here $p=\sqrt{\omega^2-\mu_i^2}$ being the momentum of the emitting  particle.  Also, $(\mu_i,g_i)$ are respectively the mass and internal degrees of freedom of the $i$-th particle species.}
%\begin{equation}
%\frac{dM}{dt} 
%=- \frac{M^4_{p}}{M^2} \sum_i g_i\epsilon_{i}\label{Page_eqn1}
%\end{equation}
%
Here, $M_p=1/\sqrt{G} \approx 1.22 \times 10^{19}\;\text{GeV} $ is the Planck mass. To the leading order approximation, the above equation can be written as:
\begin{equation}
M \simeq M_{\rm in}[1-\Gamma_{_{\rm BH}} (t-t_{\rm in })]^{\frac 1 3},
\end{equation}
with $M_{\rm in}$ being the initial mass of the black at time $t_{\rm in}$. The BH decay life time can be decoded from the above equation as,
\begin{eqnarray}\label{lifetime}
\tau_{_{\rm BH}} \simeq \frac {1}{\Gamma_{_{\rm BH}}} \simeq \frac{M_{\rm in}^3}{3 M_p^4} \,\,\frac{1}{( \sum_i g_{i}\,\epsilon_{i})}.
\end{eqnarray}
In the geometrical optical limit ($\omega \gg M_{ADM}$), %$\varphi_{s_i}(x)=1$. 
the thermal correction enters into
$\epsilon_i$ as 
\begin{equation}\label{epsilon}
    \epsilon_i= \frac{27}{8192\,\pi^{5}}\int_{z_i}^{\infty}\frac{(x^2-z^2_i)}{e^{x}- (-1)^{2s_i}}\left[1+2 \frac{(-1)^{2s_i}}{e^{y{x}}- (-1)^{2s_i}} \right]\,x \,dx
\end{equation}
%
%\begin{equation}\label{epsilon}
%\epsilon_i= \frac{27}{8192\,\pi^{5}}\int_{z_i}^{\infty}\frac{(x^2-z^2_i)}{e^{x}- (-1)^{2s_i}}\left[1+ \frac{e^{x}+ (-1)^{2s_i}}{e^{y{x}}- (-1)^{2s_i}} \right]\,x \,dx
%\end{equation} 
%
\color{black}
where, 
$x={\omega}/{T_{_{BH}}}=(8\pi M \omega / M_p^{2})$, $z_{i}=({\mu_i}/{T_{_{BH}}})$ and $y=({T_{_{BH}}/{T_b}})$, with $T_{_{BH}}$ being the temperature of black hole. In our final numerical computation we consider all the standard model particles. We limit ourselves to black holes of very small mass, and therefore with very high $T_{_{\rm BH}}$. Further, the particles emitted by the black hole is assumed to be massless ($z_i\rightarrow 0$). There are two reasons for such assumptions: First, such small mass black holes cannot be produced by any astrophysical mechanism, and hence should be of cosmological origin \cite{Carr:1974nx, Bousso:1996au,Caldwell:1996pt,Hawking:1995zn,Mann:1995vb,Khlopov:2008qy,Hawking:1994ii,Dowker:1993bt,Garriga:1999bq,Garcia-Bellido:2017mdw,Harada:2013epa}, and secondly, for these choices, the effects are much more pronounced and appreciable.\\ 

\begin{figure}[t]
    \centering    \includegraphics[width=1\linewidth]{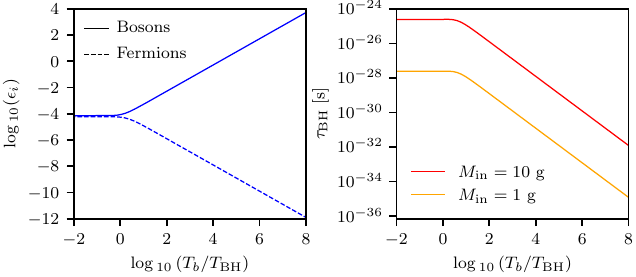}
    \caption{\textbf{Left panel :} The parameter $\epsilon_i$ is plotted as a function of $T_b/T_{\rm BH}$, as defined in equation (\ref{epsilon}), for bosons (solid line) and fermions (dashed line) in the limit $z_i \to 0$. \textbf{Right panel :} The lifetime of black holes (in seconds) is plotted as a function of $T_b/T_{\rm BH}$ for two initial black hole masses, $1 \text{g}$ (yellow) and $10 \text{g}$ (red), based on equation (\ref{lifetime}).}
    \label{fig1}
\end{figure}

The particles emitted from these small mass PBHs either can get thermalized or heat up the thermal bath depending on the bath temperature. If $T_b < T_{_{BH}}$, the emitted particles are expected to heat up the local surrounding. We ignore this scenario in our analysis (see however the recent studies in \cite{Gunn:2024xaq,Hamaide:2023ayu,He:2024wvt,He:2022wwy,Altomonte:2025hpt}). On the other if $T_b > T_{_{BH}}$, then the emitted particles are expected to get thermalized with the bath almost instantaneously. It is in this context that the thermal corrections to Hawking spectrum, eqns. \eqref{scalar_ndensity} or \eqref{fermion_ndensity}, will manifest appreciably. Our numerical analysis shows that such correction leads to an enhanced decay rate, and consequently a shorter life time of the black hole as depicted in Fig.\ref{fig1}. Indeed, the left panel of Fig.\ref{fig1} shows that in the limit of high bath temperature, as $y\rightarrow 0$, $\epsilon_i \propto y^{-1} \propto T_b$, for bosonic particles, whereas $\epsilon_{i} \rightarrow 0$ for fermionic particles. Cumulatively, such behavior of particle species translates into the following statement: the lifetime of a black hole decreases as the bath temperature increases, that is, $\tau_{_{\rm BH}} \sim 1/\Gamma_{_{\rm  BH}} \propto T_b^{-1}$. The right panel of Fig.\ref{fig1} depicts this pattern for two different values of initial PBH masses, $M_{\rm in} = (10,1)$ grams. In particular, if one puts a $10$ gram black hole in a thermal bath of temperature $T_b/T_{_{BH}} = 10^{6}$, the decay time scale of the PBH changes from $10^{-24}$ seconds to $10^{-29}$ seconds due to these corrections. \\
%%%%%%%%%%%%%%%%%%
As pointed out earlier, the cosmological background could be an ideal environment to observe the non-trivial effects arising due to a finite temperature bath. Therefore, in the following, we introduce a model system where the universe, after inflation, is assumed to be dominated by matter like field. If the spacetime is taken to be the standard Friedmann–Lemaitre–Robertson–Walker metric, $ds^2 = -dt^2 + a^2(t)(dx^2 + dy^2 +dz^2)$, for matter dominated universe the scale factor behaves as $a(t) \propto t^{2/3}$, and the Hubble parameter $H$ behaves as $H \propto a^{-3/2}$~\cite{Baumann:2018muz}. In this cosmological background the PBHs are assumed to be formed from  large curvature fluctuations, and their mass at the time of formation, $M_{\rm in}$ can be related to the Hubble constant (at that time) $H_{\rm in}$ through the relation: $M_{\rm in}\sim \gamma M_p^2/(2 H_{\rm in})$, where, $\gamma \sim 0.2$ is the collapse fraction \cite{Carr:1974nx, Carr:1975qj}. 
However, for our present purpose we shall leave the initial mass as a free parameter. 
Now, due to injection of additional entropy the temperature of thermal bath evolves differently, quantified through the relation: $T_b = T_{\rm in} (a/a_{\rm in})^{-\alpha}$, where $a_{\rm in}$ and $T_{\rm in}$ are respectively the cosmological scale factor and the bath temperature at the time of formation of PBH. Such additional entropy injection process, also known as reheating, typically occurs via decay of additional fields such as inflaton or a moduli~\cite{Kofman:1994rk,Kofman:1997yn}. For instance, if the inflaton field decays into bosonic (fermionic) radiation, the bath temperature follows $T_b \propto a^{-1/4}$ (resp. $a^{-3/4}) $ \cite{Haque:2023yra, Chakraborty:2023ocr}. When such decay process stops, and the reheating of the universe ends, the associated temperature is designated as the reheating temperature $T_{\rm re}$. After crossing $T_{\rm re}$, the universe becomes standard radiation dominated and the bath temperature evolves as $a^{-1}$, and the corresponding Hubble parameter $H$ behaves as $H \sim a^{-2}$. For model independence, we assume $\alpha \in (0,1)$. We consider a two stage reheating process with the radiation bath temperature parmeterized as $T_b = \Theta (T_b -T_{\rm eq}) a^{-\alpha_1} + \Theta(T_{\rm eq} -T_b) a^{-\alpha_2}$, with $\alpha_1=0.16,\, \alpha_2 =0.98$, and $\Theta (x)$ being the Heaviside step function.
At any instant of time, the bath temperature is defined using the Stefan-Boltzmann relation, $\rho_{R}= (\pi^{2}/30) g_{\ast} T_{b}^{4}$, where $\rho_R$ is the radiation energy density,  %The initial radiation energy density is
%\begin{eqnarray}
%    \rho_{\rm in}= \frac{\pi^2}{30} g_{\ast} T_{\rm in}^4
%\end{eqnarray}
and $g_{\ast}$ is the relativistic degrees of freedom and for our present purpose we assume its value to be $\approx 106.75$~\cite{Baumann:2018muz}. 
Note that $\alpha =1$ represents the conventional scenario when entropy is conserved. Here, the reheating process is assumed to be governed by two different decay channels, and $T_{\rm eq}$ is the temperature where these channels contribute equally. Important to note that given the initial value of the inflaton energy density, $T_{\rm eq}$ can be directly obtained from the reheating temperature $T_{\rm re}$. For our subsequent analysis we set the reheating temperature $T_{\rm re} = 10^{-2}$ GeV through out. In the following, we shall show how the decay of PBH proceeds in this model. A detailed study of this model is left for future publications.\\ 

%Further, $\alpha <1$ can arise when additional entropy is injected into the thermal bath via decay of additional field such as inflaton/any moduli. For instance, if the inflaton field decays into bosonic/fermionic radiations, is is shown that the bath temperature follows $T_b \propto a^{-1/4} / a^{-3/4} $, and if it decays into fermionic radiations, then $T_b \propto a^{-3/4}$. This continues up to $T_{\rm re}$, after which the bath temperature scales as $a^{-1}$.

\begin{figure}[t]
   \centering
   \includegraphics[width=1\linewidth]{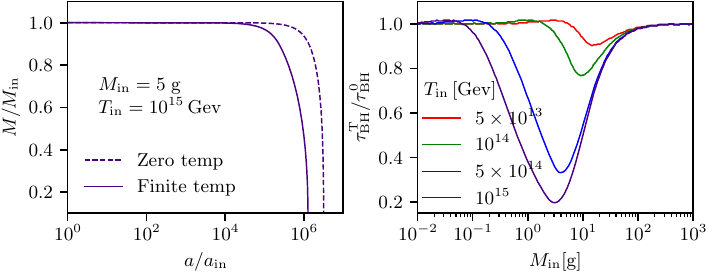}
   \caption{\textbf{Left panel :} The time evolution of normalized PBH mas $M/M_{\rm in}$ is plotted as a function of cosmological scale factor $a/a_{\rm in}$. \textbf{Right panel :} The finite temperature corrected lifetime $\tau^{\rm T}_{\rm BH}/\tau^{0}_{\rm BH}$ of a PBH is plotted as a function of its initial mass $M_{\rm in}$ for four different initial bath temperature.}
    \label{BHlife}
\end{figure}

Contrary to the previous analysis of eqn. \eqref{Page_eqn1} involving a fixed bath temperature, in our model cosmological settings, the bath temperature is evolving in time. Hence, the modified PBH evolution equation will be 
%
%0.8\pi M_p^2/H_{\rm in}
\begin{equation}\label{PBH_evolution}
\frac{dM}{da} 
=- \frac{M^4_{p}}{a H M^2} \sum_i g_i\epsilon_{i}\, .
\end{equation}
%

%This relation allow us to calculate the initial bath temperature $T_{\rm in}$ in terms of initial PBH mass as
%\begin{eqnarray}
%    T_{\rm in}= \left(\frac{\gamma}{2}\right)^{1/2} \left( \frac{4\pi^3}{45} g_{\ast} \right)^{-1/4} M_p^{3/2} M_{\rm in}^{-1/2} .
%\end{eqnarray}
We evolve this equation from matter to radiation dominated universe. To numerically solve the PBH evolution eqn. (\ref{PBH_evolution}), and scan the full parameter space, the initial PBH mass ($M_{\rm in})$ and its formation temperature $T_{\rm in}$ are kept uncorrelated, and are varied independently. %we can write the $\epsilon_i$ as a function of $a$ and also taking the contribution from all the Standard Model particles, 
It is important to note that there are two competing effects in the evolution of PBH: As the reheating progresses, whereas temperature of the thermal bath decreases, the Hawking temperature of the decaying PBH ($T_{_{BH}} \propto 1/M$ ) on the other hand increases. Such competing effects eventually singles out a particular $M_{\rm in}$ value with highest decay rate and shortest lifetime $\tau_{_{B}}$ as depicted in Fig.\ref{BHlife}. For example if we consider the PBH formation temperature $T_{\rm in} = 10^{15}$ GeV, 
the left panel of Fig.\ref{BHlife} clearly shows the finite temperature corrected decay width makes a black hole decaying earlier (solid purple line) as compared to its zero temperature counterpart (dotted purple line). The life time ($\tau_{_{\rm BH}}^{\rm T}$) of PBH of $M_{\rm in} \sim 5$ gram is maximally reduced by $10^{-1}$ from its zero temperature counterpart $\tau_{_{\rm BH}}^0$ (shown in purple color). However, with deceasing initial bath temperature ($T_{\rm in}$) the PBH would survive longer due to its reduced decay rate as expected, and depicted in different colors in Fig. \ref{BHlife}. 
To summarize, a non-extremal black hole is known to decay due to semi-classical Hawking emission which is Plankcian in nature with temperature defined though their surface gravity. Interestingly, if the said black hole is immersed in a thermal bath, and the emitted Hawking radiation is assumed to get thermalised soon thereafter, the spectrum is shown to receive a non-Planckian correction which is a function of both the bath and the black hole temperatures. If the bath temperature happens to be larger than that of the black hole, and we  neglect the effects of any ingoing classical flux or its backreaction towards the black hole dynamics, this results into an enhancement of the black hole decay rate. It is worth noting that if the energy density of the thermal bath becomes comparable to that of the BH-such as in the extreme environments of the early universe where $T_{\rm BH} \sim \sqrt{G} T_{\rm bath}^2$--the standard Schwarzschild geometry and the classical Hawking temperature relation may require modifications, as the background density can significantly alter the effective horizon and surface gravity \cite{Barrau:2025psm}. While our present analysis focuses exclusively on the modified quantum emission and the resulting enhancement of the decay rate, the overall evolution of a black hole in a thermal bath is also governed by competing accretion processes. The dynamics of black hole growth via the absorption of the surrounding thermal bath is a nuanced problem that has recently undergone critical revisions \cite{Kallifatides:2026sik, Haque:2026vvp, Das:2025vts, Kalita:2025fcs}. By treating these accretion effects as negligible in our current setup, we are able to cleanly isolate the non-Planckian corrections to the Hawking spectrum. Our preliminary studies exhibit an appreciable reduction of the life time for low mass black holes. Such low mass black holes are inevitably formed from large fluctuations in the early universe. Clearly, reduction of their life time in the cosmological environment therefore may have significant ramifications into the existing constraints on PBH cosmology. %This effect of thermalbath to the Hawking spectrum will therefore be interesting to explore futher in PBH cosmology which we defer for our future studies.
Detailed analysis is deferred for future studies.

%%%%%%%%%%%%%%%%%%%%%%%
% \begin{figure}[t]
%    \centering
%    \includegraphics[width=1\linewidth]{Two_plots_b_corrected.pdf}
%    \caption{\textbf{Left panel :} The time evolution of normalized PBH mas $M/M_{\rm in}$ is plotted as a function of cosmological scale factor $a/a_{\rm in}$. \textbf{Right panel :} The finite temperature corrected lifetime $\tau^{\rm T}_{\rm BH}/\tau^{0}_{\rm BH}$ of a PBH is plotted as a function of its initial mass $M_{\rm in}$ for four different initial bath temperature.}
%     \label{BHlife}
% \end{figure}

%
% \begin{figure}
%     \centering
%     \includegraphics[width=1\linewidth]{BH_mass_evolution.pdf}
%     \caption{Caption}
%     \label{fig:enter-label}
% \end{figure}

% \begin{figure}
%     \centering
%     \includegraphics[width=1\linewidth]{two_plots_2.pdf}
%     \caption{Caption}
%     \label{fig:enter-label}
% \end{figure}

{\bf Acknowledgments}: AC thanks the DAE- BRNS for their funding through 58/14/25/2019-BRNS. DM wishes to acknowledge support from the Science and Engineering Research Board (SERB), Department of Science and Technology (DST), Government of India (GoI), through the Core Research Grant CRG/2020/003664. JK thanks  Rajesh Karmakar for his valuable comments during the work. DM and JK also thank the HEP and Gravity group members at IIT Guwahati. \\

\end{document}